# Salud pública y el acceso a los medicamentos:

[El papel de la industria farmacéutica]

Universidad Loyola Andalucía.

*Trabajo dirigido por José Vallés Ferrer.*

*Juan B. González Blanco, Economía y Relaciones Internacionales*

# Índice

# 1. Introducción

Según la Organización Mundial de la Salud, cada hora que pasa mueren en el mundo unas 1.150 personas por falta de acceso a medicamentos para enfermedades curables. Esto supone una lacra de 10 millones de personas cada año, muertes perfectamente evitables con el actual desarrollo de la medicina moderna. Al mismo tiempo, en Estados Unidos mueren cada año 128.000 personas por tomar medicamentos que, aunque se los recetara la médica, no eran seguros. Pueden parecer dos problemas nada relacionados, pero no son más que dos caras de la misma moneda, que se refleja de forma distinta en los países desarrollados y los que están en desarrollo.

Es este un problema al que no se le presta ninguna atención en medios de comunicación ni en debates políticos. Por tanto, el objetivo de este trabajo será arrojar luz sobre estos hechos, analizar sus múltiples causas y tratar de ofrecer soluciones basándonos en modelos de economía de la salud pública exitosos.

Multitud de hipótesis han sido formuladas para hallar las causas de tan acuciante situación. La falta de instituciones consolidadas en los países en desarrollo, la corrupción, la dejadez de los servicios públicos de salud, etc., han sido algunas de las respuestas que se han ofrecido, en ocasiones desde investigaciones alineadas con la industria farmacéutica. Estas causas, sin dejar de ser ciertas en los países en desarrollo, no pueden ser aplicadas a países altamente desarrollados como los Estados Unidos o los países de la Unión Europea, donde también se producen casos de falta de acceso a medicamentos.

Por tanto, la OMS reconoce que hay otros factores aún más determinantes para la falta de acceso a los medicamentos esenciales: el bajo poder adquisitivo de los enfermos y el alto precio de los medicamentos disponibles (WHO, 2006). En estos factores nos centraremos a la hora de analizar la situación, pero también la veremos de una manera más holística para poder presentar algunas propuestas que contribuyan a mejorar la salud de las personas. Por supuesto, el problema del bajo poder adquisitivo de los enfermos en países en vías de desarrollo no es algo nuevo, y se ha abordado el tema desde muchos puntos de vista, con gran multitud de propuestas orientadas a acabar con este problema. Por esto, nos centraremos más en la formación de los precios de los medicamentos y qué papel juegan en ella las grandes compañías de la industria farmacéutica, las llamadas *Big Pharma*.

Por otro lado, las enfermedades y las muertes en los países desarrollados por medicamentos que no han sido suficientemente estudiados (recordemos el caso de la talidomida) son a menudo casi tan olvidados por el mundo como aquellos que mueren sin poder acceder a un tratamiento adecuado en la otra punta del globo. De este tema no se ha escrito tanto, y se ha investigado bastante poco, por lo que no disponemos de demasiados datos fiables a día de hoy. Pese a esta falta de información y de fuentes fiables, es necesario mencionar que la reacción adversa a los medicamentos, como luego veremos, se ha convertido en la cuarta causa de muerte en países tan desarrollados como Estados Unidos, por encima de los accidentes de tráfico o los homicidios, que tanta atención mediática reciben.

Una problemática tan multicausal como la que nos ocupa requiere que la abordemos por partes, para poder abarcarla en toda su complejidad. Por ello, comenzaremos por ver la situación del acceso a los medicamentos en el mundo, concentrándonos especialmente en los países en desarrollo, que, como veremos, es donde más se sufre esta lacra. Analizaremos las causas más importantes de este problema que afecta a millones de personas en todo el mundo, y veremos más de cerca algunos de los casos reales que se han producido a lo largo de la historia por acción u omisión de la industria farmacéutica, especialmente en la región del África Subsahariana.

Tras este primer análisis de las causas de la falta de acceso, haciendo especial énfasis en el papel que juegan los derechos de propiedad intelectual en la formación de los precios de los fármacos, nos moveremos hasta los países más desarrollados, sede de las grandes productoras farmacéuticas. En ellos podremos estudiar el comportamiento de la industria farmacéutica con respecto a los países en vías de desarrollo, las relaciones de poder e influencia que se establecen entre estas empresas y los gobiernos, y qué papel juegan en la tragedia de la falta de acceso a los medicamentos.

Ello nos llevará inevitablemente a un análisis del funcionamiento de esta industria no sólo en sus relaciones con el extranjero, sino en los propios países donde sus sedes están establecidas. Profundizaremos en los argumentos que la industria utiliza para defender una mayor protección a la propiedad intelectual, y descubriremos cómo funciona el sistema de investigación y desarrollo que lleva a cabo la industria. Veremos la relación que hay entre la falta de acceso de medicamentos en el Sur global y las muertes por medicación inadecuada en el Norte, y qué responsabilidad tienen en todo esto las instituciones del Estado y las grandes empresas.

Finalmente, esbozaremos algunas de las propuestas que se han planteado para abordar una situación tan compleja como esta, estudiando más de cerca uno de los casos de política económica más exitosa para la salud pública según la Organización Mundial de la Salud.

Puede pensarse que la salud pública no es un campo de la política económica clásica, y que poco pueden hacer los mecanismos más conocidos de la política económica, como la política monetaria o la política fiscal, para asegurar la salud de las personas. Sin embargo, como hemos visto, la investigación y la producción de fármacos está desequilibrada. Mientras en una parte del mundo salen al mercado medicamentos nuevos constantemente, las enfermedades que asolan otras regiones son olvidadas y deliberadamente ignoradas. El sistema de mercado deja de producir medicamentos para enfermedades peligrosas pero poco rentables hasta que no se descubre un uso cosmético para esos medicamentos. Podríamos seguir así eternamente, poniendo ejemplos de cómo los recursos humanos y los fondos de investigación, por no hablar del capital fijo de producción de las empresas, se usan en busca de beneficio económico en vez de para lograr satisfacer las necesidades humanas más básicas, como es la salud.

Como vemos, todo ello encaja con la famosa definición de economía de Lionel Robbins, "la economía es la ciencia que estudia el comportamiento humano como una relación entre fines y medios escasos susceptibles de usos alternativos" (Robbins, 1932), o con la definición de economía según la Real Academia Española: "Ciencia que estudia los métodos más eficaces para satisfacer las necesidades humanas materiales, mediante el empleo de bienes escasos" (RAE, 2018). De ahí precisamente la pertinencia de hacer un análisis desde la política económica de cómo asignar eficazmente unos recursos tan escasos como el capital humano y los fondos de investigación, además de cómo establecer la regulación adecuada sobre el mercado, con el objetivo principal y prioritario de garantizar el acceso a la salud.

Por todo ello, a lo largo de este trabajo nos encontraremos con diversas situaciones susceptibles de una "reasignación" más eficaz de los recursos, y otras donde directamente hay que empezar a construir desde cero. Todas las abordaremos con el mismo objetivo en mente: utilizar las herramientas de las que disponen los Estados para poder garantizar unos de los más básicos derechos humanos: el derecho a la vida y a la salud.

## 2. El acceso a los medicamentos en el mundo

Según la Organización Mundial de la Salud, la tercera parte de la población mundial no tiene acceso a medicamentos esenciales para su salud. Es decir, en pleno siglo XXI, más de 2 mil millones de seres humanos ven en peligro su salud y en riesgo su vida por no poder comprar los medicamentos que necesitan. Por supuesto, siguiendo los patrones de desigualdad material cada vez más consolidados entre las economías del Norte y del Sur globales, los países en desarrollo son los que más sufren este problema. Si bien en el mundo 1 de cada 3 personas carece de acceso, este número se eleva a 2 de cada 3 en los países en desarrollo (Lage Dávila, 2011).

La dificultad para acceder a medicamentos esenciales nos puede parecer algo muy lejano, especialmente en un país con un sistema sanitario público como España. Sin embargo, según el Ministerio de Sanidad, Consumo y Bienestar, en 2017 el 4,7% de la población española, unos 2,2 millones de personas, no pudieron acceder a los medicamentos recetados por su médico por problemas económicos (MSCB, 2018). ¡Y esto en un país en el que la sanidad pública está garantizada para todos los ciudadanos! Si nos vamos a los Estados Unidos de América, este número asciende hasta un 16,8% en los Estados Unidos para los mayores de 55 años (aquellos que pueden optar a los programas públicos de medicamentos), unos 55 millones de personas (Morgan & Lee, 2017). El gasto en salud ha ido creciendo para todas las familias en los países desarrollados. Por ejemplo, mientras que una trabajadora que gane el sueldo mínimo en California en 1970 aseguraba la salud a su familia de 4 miembros con el 15% de su sueldo anual, en 2005 la misma cobertura le costaría el 106% del salario (Lage Dávila, 2011). Este es un claro efecto de las políticas neoliberales que se aplicaron a partir de los años 70 en todo el mundo, pero, por motivos de extensión, no podemos detenernos a estudiar este fenómeno.

Por tanto, si los enfermos no pueden acceder a los medicamentos que necesitan en los países más "ricos", ¿cuál es la situación en aquellos países donde la población dispone de menos recursos?

### 2.1. Las enfermedades olvidadas

Hemos visto que 2 de cada 3 personas en los países en desarrollo no pueden disponer de medicamentos esenciales para su salud. ¿Por qué sucede esto? En primer lugar, podemos destacar que hay enfermedades que sólo afectan a estos países, normalmente enfermedades tropicales, a las que se destina una ínfima parte de los recursos de Investigación y Desarrollo (I+D). Algunas de estas son conocidas como *enfermedades tropicales olvidadas*, porque pese a causar más de medio millón de muertes cada año, los esfuerzos para combatirlas son muy reducidos (Hotez &

Sachs, 2007). Son la sexta causa de pérdida de años de vida (indicador DALY) en el mundo, la segunda si incluimos la tuberculosis, y sin embargo de todos los fármacos descubiertos entre 1980 y 2004 sólo el 1% iba destinado a estas enfermedades (Lage Dávila, 2011).

Como podemos ver, bien merecen el nombre de enfermedades olvidadas. La razón por la que esto ocurre es porque la mayor parte de las nuevas medicinas las desarrollan empresas privadas que se mueven por puro interés económico. Las grandes compañías farmacéuticas analizan costes y beneficios potenciales de cada fármaco en desarrollo, y comparan esos beneficios potenciales con los que podrían obtener invirtiendo ese mismo volumen de capital en el mercado bursátil a un rendimiento del 11% (Light & Warburton, 2011). Por tanto, una medicina no se investiga a si su margen de beneficio se estime menor que esas cifras.

Con esta ética, en la que el medicamento es considerado un bien comercial y no un bien social, se espera que el mercado abastezca a una población con pocos recursos. Por supuesto, las enfermedades que afectan a las zonas menos desarrolladas del planeta son ignoradas por la baja previsión de beneficios. No importa que el hecho de no investigar un nuevo fármaco conlleve miles de muertes innecesarias.

Puede parecer chocante que empresas que afirman que su misión es "descubrir nuevas maneras de mejorar y prolongar la vida de las personas" [1] se olviden de un gran número de personas a la hora de investigar. Veamos algunos ejemplos para aclararlo.

La infección de ceguera de los ríos (*oncocercosis*) es la segunda causa de ceguera a nivel mundial, con una prevalencia estimada de 127 millones de infectados ("WHO | Onchocerciasis - river blindness," 2010) y que va en aumento. De ellos, el 96% vive en África. Su nombre proviene de la mayor concentración del agente infeccioso en las zonas cercanas a los ríos, por lo que muchos abandonan estas tierras (las más fértiles) por miedo a la enfermedad. Todo ello supone un gran lastre para el desarrollo de amplias zonas de los países africanos. Hasta finales de los años 70, las grandes farmacéuticas no habían destinado recursos para la investigación de esta enfermedad, cuando por casualidad, habiendo desarrollado un antibiótico para uso veterinario (ivermectina), Merck descubrió que podía servir también para la ceguera de los ríos. Este es, 40 años después, el tratamiento más eficaz conocido frente a la segunda causa de ceguera en el mundo: una serendipia (WHO, 2006).

Otro ejemplo es el caso de la *Tripanosomiasis* africana, comúnmente llamada enfermedad del sueño. Durante la segunda mitad del siglo XX, esta enfermedad se trataba con melarsoprol, un derivado del arsénico, por lo que el tratamiento también era altamente peligroso. Finalmente, ante las grandes epidemias que se cebaban con algunos países africanos que la convirtieron en una causa de muerte aún más común que el SIDA en ciertas zonas ("WHO | Trypanosomiasis," 2012), se descubrió un tratamiento mucho más eficaz y seguro, la eflornitina. Sin embargo, en 1995, cuando aún quedaban cientos de miles de personas infectadas, las compañías farmacéuticas abandonaron la producción del fármaco por falta de rentabilidad. La producción no se retomó hasta años después, cuando se descubrió un uso cosmético de la eflornitina (depilación facial) que sí era rentable pues lo adquirirían mercados de altos ingresos (WHO, 2006).

---

[1] Novartis: https://wsww.novartis.es/sobre-novartis/quienes-somos/mision

Así, podemos afirmar que las empresas farmacéuticas no son más que eso, empresas, y se mueven en busca de beneficios económicos. Si beneficios de salud acompañan a esas rentas dinerarias es una preocupación secundaria, como más adelante veremos. Por ello, las enfermedades que la Organización Mundial de la Salud califica como de Tipo III (las que afectan sólo a países en desarrollo) no reciben la atención necesaria y muchas veces ni siquiera existen tratamientos efectivos para frenar su avance. Esta es una de las causas de falta de acceso a los medicamentos.

Pero, ¿qué ocurre cuándo sí que existen tratamientos efectivos para frenar el contagio de una enfermedad y tratar sus síntomas? En este caso la perspectiva no mejora especialmente para los enfermos en las regiones menos desarrolladas del planeta. Pongamos el caso del SIDA, una enfermedad ya muy estudiada, con tratamientos antirretrovirales (ARV) muy eficaces. Pese a ello, en algunos países del África Subsahariana como Botswana, Lesoto, Suazilandia o Zimbabue la esperanza de vida al nacer ha decaído en unos 20 años desde 1980 (Torres Domínguez, 2010). Para poder ofrecer una respuesta ante estos hechos analizaremos las razones de la falta de acceso a medicamentos en los países en desarrollo.

## 2.2. Las razones de la falta de acceso: el poder adquisitivo

La Organización Mundial de la Salud sentó cátedra en su conocido informe sobre salud pública, innovación y derechos de propiedad intelectual de 2006: "El precio de las medicinas y otros productos de salud, incluso cuando se venden a 'precios de coste' en los entornos más pobres, y la capacidad para pagar por ellos, son los factores críticos para ampliar o restringir su acceso" [2]. Así, nos centraremos en el abismo existente entre los precios de los medicamentos y el poder adquisitivo de los enfermos en los países en desarrollo.

En primer lugar, concretar el marco estadístico en el que nos movemos. ¿Cuándo se considera que un medicamento tiene un precio superior al que se pueda permitir un enfermo? Siguiendo de nuevo a la OMS, en este caso a su informe decenal sobre la situación de los medicamentos en el mundo (el último disponible es de 2011), un tratamiento se considera demasiado caro si cuesta más que un día de salario del sueldo más bajo que cobren los funcionarios del país. Este método, aunque cuenta con la ventaja de ofrecer una medida homogénea y comparable, tiende a sobrevalorar la accesibilidad de un medicamento, ya que un número incierto de personas trabajan cobrando menos que el mínimo de las administraciones públicas. Por eso, aunque según esta metodología una medicina sea accesible, esto no quiere decir que no haya un alto número de personas que no tengan una renta suficiente para adquirir el tratamiento. Hay que tener en cuenta esta subestimación en el número de personas privadas de acceso según los datos de la OMS.

---

[2] Traducción propia. Original: *The price of medicines and other health products, even when "at cost" in the poorest settings, and the ability to pay for them, are the critical factors in enhancing or hindering access.* (WHO, 2006: 109)

Mediante el estudio de encuestas en países representativos de diferentes regiones en desarrollo del mundo (Latinoamérica, África, Sudeste asiático y Oceanía) obtuvo la OMS los datos de la capacidad de compra de medicamentos. Veamos los datos del tratamiento contra las infecciones respiratorias, las enfermedades más comunes en el mundo (Allen, 2016). En ningún país de los estudiados, un funcionario cobrando el sueldo mínimo de la administración podría pagarse el tratamiento recomendado para la infección si se usasen medicamentos originales ("de marca"), y sólo en 5 países sería posible si se usasen medicamentos genéricos [3] (WHO, 2011).

Lo mismo ocurre con la diabetes: sólo en 2 de los países estudiados era accesible el tratamiento, incluso utilizando medicamentos genéricos. Cabe destacar que entre estos países "en desarrollo" se incluyen los Emiratos Árabes Unidos, con un PIB per cápita de 40.400 dólares en 2011 (WB, 2018), el año que se presenta el informe. Este no es el caso evidentemente del resto de países en desarrollo, especialmente del África Subsahariana, así que debemos sumar a la subestimación del problema según la OMS.

Por tanto, podemos ver que la falta de acceso a medicamentos por los altos precios de estos es un problema que afecta a gran parte de las personas que viven en países en desarrollo. Este drama se acrecienta conforme las enfermedades sean más duraderas: si ya es complicado poder adquirir un tratamiento de una semana para curar una infección respiratoria, es prácticamente imposible adquirir el tratamiento para una enfermedad crónica que requiera de medicación constante durante años. Por tanto, ante esta limitación, debemos recordar que las personas más vulnerables son aquellas que sufren enfermedades crónicas.

Esta falta de acceso se ve aún más condicionada por el hecho de que en los países en desarrollo, los pacientes deben pagarse sus propias medicinas en la inmensa mayoría de los casos: hasta un 90% de la población tiene que recurrir a pagarse los tratamientos "de su propio bolsillo" [4] (WHO, 2011). Esto puede suponer un gasto inmenso en regiones de bajas rentas, como podemos ver en el Mapa 1 en el Anexo Cartográfico, por lo que estos gastos de su propio bolsillo empujan a muchas familias al endeudamiento, a tener que escoger entre comer o pagar los tratamientos necesarios. (*op. cit.*). De hecho, según las cifras del Banco Mundial [5], cada año entre 1995 y 2014, unas 388.400 personas se han visto empujadas por debajo del umbral de pobreza (3,10$ PPA), de media, en cada país del África Subsahariana. Este empobrecimiento y endeudamiento de las familias tiene otros efectos a largo plazo, principalmente la falta de capacidad para poder pagar tratamientos futuros (como hemos visto, este problema crece exponencialmente en el caso de las enfermedades crónicas) y la reducción de otros factores de salud, como una nutrición adecuada. De esta manera, la enfermedad y falta de acceso a cuidados empuja a los habitantes de los países en desarrollo a la pobreza, y esto a su vez hace que empeore su salud a largo plazo, y así sucesivamente. Este círculo vicioso entre pobreza y salud se conoce como "trampa de la pobreza", una trampa de la que es casi imposible escapar por uno mismo, ya que la mala salud conlleva una peor productividad y menos posibilidades de obtener las rentas del trabajo suficientes como para saldar deudas y tener la capacidad de pagar los tratamientos (Perry, Arias, López, Maloney, & Servén, 2006).

---

[3] Un medicamento genérico es aquel que no es producido por el laboratorio que desarrolló el principio activo (el que lo produjo por primera vez y lo patentó) (González, Fitzgerald, & Bermúdez, 2006).

La falta de desarrollo de los sistemas públicos de salud tiene bastante que ver con que las familias tengan que destinar grandes proporciones de su renta a comprar medicamentos de su propio bolsillo. Esta carencia administrativa estatal es demasiado compleja como para ser abordada en un estudio de esta extensión, pero resulta imprescindible presentar algunos datos que puedan servir de reflexión.

A finales de los años 80, el Fondo Monetario Internacional, el Banco Mundial, los países más desarrollados (liderados por los Estados Unidos) y los representantes de los grandes bancos privados diseñaron un plan de política económica basado en las ideas neoliberales, al que se denominó "Consenso de Washington" (Williamson, 1989). Este plan, que se pensó especialmente para los países de América Latina, pronto se impuso a nivel global como requisito para poder recibir préstamos internacionales e incluso ayuda de los países desarrollados. África, debido a la baja capacidad recaudatoria de los Estados en la mayoría de los casos, se vio obligada a aceptar la liberalización de sus recursos y la reducción de los presupuestos sociales para priorizar el pago de la deuda externa sobre el bienestar de los ciudadanos (Vallés Ferrer & Chaban, 2015).

---

[4] En la bibliografía especializada en inglés, el gasto en medicinas no cubierto por los sistemas sanitarios públicos o los seguros de salud se denomina *Out Of Pocket expenditure*.

[5] Cálculos del autor a partir de los datos de los *World Development Indicators* (WB, 2018). Hay que tener precaución a la hora de analizar estas cifras ya que en la base de datos del Banco Mundial, pese a ser la más completa del mundo, los datos que se recogen sobre gastos *Out Of Pocket* son bastante dispersos en el tiempo y no de todos los países. Por tanto, es sensato considerar los guarismos presentados como una estimación que, como hemos visto con los datos de la OMS, esté infravalorada.

Como consecuencia de todo ello, los 46 Estados del África Subsahariana destinaron de media, entre los años 1995 y 2014, un 64% más en servicio de la deuda externa (devolución de la deuda, intereses y refinanciación) que a presupuesto de sanidad pública. Algunos años llegaron a pagar un 252% más en beneficio de los países desarrollados y fondos de inversión que les prestaron el dinero que en la salud de su propia población [6]. Cabe recordar que esta priorización de la deuda es una política económica impuesta desde el exterior, como mecanismo para asegurar a los acreedores internacionales la devolución de la deuda y de sus intereses (los principales acreedores de la deuda externa son grandes bancos y fondos de inversión de los países desarrollados (Toussaint, 2016)).

Los bajos presupuestos en sanidad tienen graves consecuencias en la política sanitaria. Los altos precios a los que se ven obligados los Estados a comprar los medicamentos (como veremos más adelante) hace que la partida de compra de fármacos suponga entre un 25 y un 75% de los presupuestos en sanidad en los países de rentas bajas y medias, mientras que en los países de rentas altas supone sólo un 15% (Lage Dávila, 2011). Esta distribución hace que las partidas destinadas a infraestructuras y a capital humano tengan que reducirse, haciendo que las condiciones para los trabajadores sanitarios y los pacientes no sean las adecuadas. Además, los bajos sueldos y difíciles condiciones de trabajo que ofrecen las administraciones públicas de los países en desarrollo han provocado una hemorragia de profesionales sanitarios. Por ejemplo, de los médicos que iniciaron su carrera en los años 90, en 2005 ya habían emigrado de Zimbabue un 70% y de Ghana un 60%. Debido a esta hemorragia, se estima que para poder cubrir las necesidades de la región del África Subsahariana, el número de profesionales sanitarios debería triplicarse (WHO, 2006). Estas prioridades entre deuda y bienestar no sólo se notan en el gasto en sanidad, también provocan que se resienta el sistema educativo que, como veremos posteriormente, es uno de los pilares básicos de una política de salud pública efectiva.

Por suerte, se aprecia una tendencia al alza en el presupuesto destinado a sanidad en estos países durante los últimos años, y el servicio a la deuda se va reduciendo lentamente con el paso del tiempo. La imposición exterior de la prioridad de la deuda frente a la salud pública supone una violación de la soberanía de estos Estados. Como dijo el Relator Especial de ONU para el Derecho a la Alimentación Jean Ziegler, "el servicio a la deuda es un gesto visible de sumisión" (Ziegler, 2005). Debemos aprender muchas lecciones de los fracasos pasados, y en el aspecto social, el Consenso de Washington fue claramente uno de ellos (Vallés Ferrer & Chaban, 2015).

---

[6] Cálculos del autor en base a los datos de los World Development Indicators (WB, 2018).

Recapitulando, hemos visto cómo las familias en los países en desarrollo, especialmente en África Subsahariana, no pueden acceder a los medicamentos porque los servicios públicos de salud no son capaces de ofrecerlos a precios más bajos, por lo que la mayoría se ve obligada a comprarlos de su propio bolsillo, lo que supone un gran esfuerzo económico, a veces insuperable, para los hogares de rentas bajas y medias. Pero en todo momento hemos considerado que los precios son muy altos como algo exógeno, sin llegar a preguntarnos siquiera por qué. ¿Es que hay alguna razón para unos precios tan astronómicos comparados con el poder adquisitivo de los mercados en desarrollo?

## 2.3. Razones de la falta de acceso: los precios

El bajo poder adquisitivo de las familias no es el único factor que restringe su acceso a los medicamentos. Si las empresas farmacéuticas que producen los tratamientos hicieran una discriminación por precios entre sus mercados, de manera que vendieran más baratos sus productos en los países con menores rentas, se podría superar en parte la barrera del poder adquisitivo. Sin embargo, si vendiesen sus productos a precios más bajos en el Sudeste Asiático que en Europa, las empresas se arriesgan a que se produzcan "importaciones paralelas" [7] o que esos precios se conviertan en los de referencia internacional para las negociaciones con los Estados de regiones más ricas (que suponen grandes contratos de compra para las farmacéuticas) (WHO, 2011). Por ello, los precios de los medicamentos patentados tienden a homogeneizarse a nivel global, pero tendiendo hacia los precios más altos de Estados Unidos y Europa (Torres Domínguez, 2010). De esta forma, las zonas más pobres del planeta tienen que pagar más con respecto a su renta que las zonas más ricas.

Por supuesto, también se puede recurrir al mercado de medicamentos genéricos. Los medicamentos genéricos son siempre más baratos que los medicamentos patentados (se estima que unas 3 veces más baratos como mínimo para países en desarrollo (Lage Dávila, 2011)), por lo que serían más fáciles de adquirir para la población de estas regiones. Sin embargo, la falta de desarrollo institucional implica bajos controles de calidad de los productos farmacéuticos, por lo que en muchos casos los medicamentos de precios más bajos a los que se tiene acceso son o bien falsificaciones sin ningún efecto médico, o tratamientos ya "superados" por el desarrollo de la ciencia médica. Por tanto, se estima que algo más del 25% de las familias de países en desarrollo consumen fármacos de nula o baja calidad debido a que no pueden pagar otros tratamientos más caros (WHO, 2006).

---

[7] La importación paralela es la importación de un producto adquirible en el territorio nacional, producido por la misma compañía y de la misma calidad, debido a que en el extranjero es más barato ese mismo producto.

Además, los medicamentos genéricos son en su mayoría productos importados desde la India y China, y por tanto los precios tienden a subir por las tarifas portuarias, aranceles e impuestos estatales, pero sobre todo por los márgenes de beneficios (*mark-up*) que imponen los importadores y vendedores al por mayor dentro de los propios países. La Organización Mundial de la Salud calcula que, para medicamentos genéricos de bajo coste, los impuestos estatales al consumo y a la importación suponen hasta un 55% del precio, mientras que los beneficios hacen que el precio ascienda sobre el de coste un 123% en Nigeria o un 246% en Ghana, por ejemplo (WHO, 2011). Pero, sobre todos estos factores, se impone la gran dificultad que tienen los genéricos para poder entrar en los mercados.

La entrada de medicamentos genéricos, siempre más baratos, en estos mercados, podría suponer una diferencia fundamental en la competencia siempre y cuando se apliquen controles de calidad que aseguren que son seguros y efectivos. Este aumento de la competencia haría bajar los precios de todos los medicamentos con competidores, incluso los originales, y por ello se considera la política de desarrollo de genéricos la más efectiva para poder mejorar el acceso a tratamientos sanitarios en los países en desarrollo (Tobar & Sánchez, 2005).

Algunos países del África Subsahariana han desarrollado planes de industrialización farmacéutica con el objetivo de crear una producción local para cubrir el suministro de ciertos medicamentos genéricos en el país (UNIDO, 2015). Pese a esto, la producción nacional de medicamentos, sin establecer estrategias efectivas de transferencia tecnológica y de cooperación tanto en educación superior como en investigación y desarrollo, no será competitiva frente a otros modelos de producción farmacéutica (Osewe & Nkrumah, 2008). Como veremos más adelante, la producción regional de genéricos es la respuesta a las limitaciones que imponen las fronteras nacionales, permitiendo el intercambio tecnológico y técnico necesario para hacer la industria farmacéutica más competitiva y al servicio de la salud pública.

Hemos visto que uno de los factores que limita la disponibilidad de los medicamentos genéricos es la dificultad a la que se enfrenta esta industria para acceder a los mercados, ya sean desarrollados o en desarrollo. Pero, ¿cómo se articula esta dificultad? ¿Quién tiene intereses en que no se produzcan los fármacos que podrían salvar la vida a miles de personas?

# 3. La industria farmacéutica global

En primer lugar, sería interesante presentar un esquema que resuma la estructuración hacia la que tiende la industria farmacéutica a nivel global:

**Tabla 1**. Clasificación de la industria farmacéutica según Haakonsson.

| Segmento | Función | Poder dentro de la cadena | Ejemplo |
|---|---|---|---|
| Productores con marca | · I+D <br> · Control de las patentes clave y reputación | Gobernanza | Pfizer |
| Productores genéricos de calidad | · Producción en masa para grandes mercados | Dependencia de I+D externo | India |
| Productores genéricos de bajo valor | · Producción para cubrir los nichos más pobres del mercado. | Nulo | Productores no controlados del Tercer Mundo |

Fuente: elaboración propia a partir de Haakonsson, 2009

Como podemos observar, la industria farmacéutica está claramente dividida en 3 ramas que operan con diferentes funciones y con distintas estrategias. En este apartado nos centraremos en el segmento dirigente de la cadena global de valor: las grandes empresas farmacéuticas que controlan los activos clave y ejercen la gobernanza a lo largo de la cadena, las llamadas *Big Pharma*.

Estas empresas son aquellas que tienen capacidad para fijar los precios en el mercado a través de los monopolios que ejercen sobre ciertos productos a través del sistema de patentes. Son estas las empresas que limitan el acceso a genéricos por proteger sus derechos de propiedad intelectual, y las que marcan el camino que debe seguir la investigación y el desarrollo en todo el mundo.

## 3.1. El sistema actual de patentes

La regulación de las patentes y la protección de los derechos de propiedad intelectual ha sido durante el siglo XX materia de acción de cada Estado, estando bajo su soberanía el sistema regulatorio nacional. Sin embargo, durante las últimas décadas el sistema de protección intelectual se ha ido globalizando, homogeneizándose por todo el mundo. En el año 1994 se firma el Acuerdo sobre los Aspectos de los Derechos de Propiedad Intelectual relacionados con el Comercio (ADPIC, pero nos referiremos a él por sus siglas en inglés: TRIPS), una condición obligatoria para poder ser miembro de la Organización Mundial del Comercio y de la globalización económica que se estaba intensificando a finales del siglo XX.

En los Acuerdos TRIPS se impuso un plazo de 10 años a todos los países para hacer converger las regulaciones nacionales al alto nivel de protección diseñado por las naciones más desarrolladas. En resumen, se ampliaba el plazo de protección a un producto patentado a 20 años desde la solicitud (art. 33), y se privaba a cualquier tercer actor de producir sin el consentimiento del dueño de la patente (art. 28). Los criterios de patentabilidad se hacían uniformes y, por tanto, una empresa podía obtener el monopolio a nivel mundial de la producción del invento patentado (WTO, 1994).

Pese al alto nivel de protección impuesto, también se incluyeron en el acuerdo lo que se llaman "flexibilidades" para la propiedad intelectual (art. 31). De estas, la más importante es la posibilidad de los Estados de emitir *compulsory licences* (licencias obligatorias), que permite a un gobierno autorizar la producción de un fármaco (en el caso que nos ocupa) patentado en caso de emergencia de salud pública. Se podría suponer que esto evita el problema de falta de acceso a medicamentos patentados: cuando sea necesario para mejorar la salud de los ciudadanos, un Estado puede emitir una licencia y empezar a producir el fármaco necesario. Por supuesto, no es tan fácil, ya que los acuerdos TRIPS fueron diseñados para proteger los intereses de las grandes empresas farmacéuticas (WHO, 2006). Básicamente, si un Estado no tiene desarrollada una industria farmacéutica previamente, es inútil que emita una licencia obligatoria ya que no será capaz de producir el fármaco. Esto evidentemente supone una gran limitación para los países en desarrollo, que no tienen en la mayoría de los casos una industria local lo suficientemente preparada. Además, para poder emitir la licencia hay que cumplir ciertas condiciones, por ejemplo, que se haya solicitado una licencia voluntaria previamente a la dueña de la patente y que esta la haya denegado durante un plazo prudencial (art. 31). Cabe recordar que la empresa simplemente puede posponer la decisión durante ese plazo prudencial (algo muy flexible), y tras ello frenar la implementación de la licencia mediante litigios y demandas durante un tiempo (como veremos posteriormente).

Siendo esto así, las flexibilidades incluidas en el Acuerdo son claramente insuficientes para la salud pública de muchos países en desarrollo. Pero la *Big Pharma* no se contenta con ello. Para limitar la posibilidad de que se implementen esas flexibilidades, las empresas farmacéuticas presionan a sus Estados [8] para que en los acuerdos bilaterales de comercio introduzcan medidas que incrementen aún más la protección a los derechos de propiedad intelectual de sus empresas (WHO, 2006). Podríamos pensar que esto beneficia la innovación en todas las regiones por igual, pero hay que tener en cuenta la gran asimetría que se produce en el mercado de la propiedad intelectual: mientras que los países en desarrollo producen alrededor de la tercera parte de los *papers* científicos del mundo, sólo tienen el 4,5% de las patentes (Lage Dávila, 2011). Si miramos las dos últimas décadas (desde 1997), la Unión Europea registró 10 veces más patentes que África, y los países con rentas altas 111 veces más que los de rentas bajas (WIPO, 2018). En el Mapa 2 del Anexo podemos observar la inmensa diferencia que hay entre las regiones del mundo en materia de patentes desde la aprobación de los Acuerdos TRIPS. Como se puede deducir a partir del Mapa, una mayor protección a los derechos de propiedad intelectual no ha servido para impulsar la innovación en los países en desarrollo, lo que provoca que las asimetrías entre países desarrollados y en desarrollo vaya aumentando a nivel de investigación. A esta misma conclusión llegó el Informe sobre Propiedad Intelectual de la Organización Mundial de la Salud: no hay evidencias que respalden el argumento de que las patentes sirvan para fomentar el desarrollo de nuevos productos en los países que más lo necesitan (WHO, 2006).

---

[8] De las 10 con mayor volumen de ventas en 2016, que controlan el 43,4% del mercado global (EvaluatePharma, 2016), 6 tenían su sede en los Estados Unidos, 2 en Suiza y 2 en la Unión Europea (Juncal, 2016).

De esta forma, una protección más restrictiva de estos derechos beneficia claramente a las grandes empresas que controlan las cadenas de valor globales y el mayor número de patentes. ¿Cómo logran semejante influencia? A través de lo que se conoce como *lobbying*, es decir, destinar ingentes cantidades de dinero a convencer a los políticos y a los asesores que les orientan: en la última década y sólo en Estados Unidos, la industria farmacéutica destinó más de dos mil millones y medio de dólares en *lobbying* a políticos (McGreal, 2017). Se estima que sólo en Estados Unidos, las *Big Pharma* tienen en nómina a 2 *lobbyists* por cada congresista (Forcades i Vila, 2006).

Dos mil millones y medio de dólares sólo en conseguir influir en la legislación para convertirla en normas a favor de sus propios intereses. Tienen tal poder de influencia que consiguen que sus Estados se conviertan en sus abogados personales para defender sus intereses en los acuerdos comerciales. Para hacernos una idea del inmenso poder que tienen estas empresas, veamos el Mapa 3 en el Anexo Cartográfico, que muestra el Producto Interior Bruto de los países de África comparados con los ingresos en 2016 de la empresa farmacéutica más grande del mundo, la firma norteamericana Pfizer.

Como se puede observar, sólo 9 de los 52 países tienen más PIB que ingresos Pfizer. Si repetimos la comparación, pero esta vez usando los beneficios netos tras impuestos de Pfizer el último año, obtenemos que de nuevo el PIB de sólo 12 países del África Subsahariana los supera. Teniendo semejante poder económico la mayor empresa del mundo (recordemos que "sólo" representa el 6% del mercado (EvaluatePharma, 2016)), imaginemos lo que puede conseguir la industria al completo cuando se amenazan sus intereses. Aunque en realidad no hace falta imaginarlo, sólo hay que estudiar el caso de Sudáfrica en los últimos años de la década de 1990.

En 1994 Sudáfrica ya había transpuesto algunas de las exigencias de los Acuerdos TRIPS en sus leyes nacionales. Eran los peores años del SIDA: en Sudáfrica morían por la enfermedad unas 160 mil personas cada año, y alrededor del 15% de la población adulta la sufría (GHO, 2018). El culpable: la incapacidad del sistema sanitario para adquirir los medicamentos antirretrovirales necesarios por su alto coste, ya que estaban patentados (además de la regulación existente en el momento). Es de esta época la denuncia de Eric Goemaere, coordinador de la campaña contra el SIDA en Sudáfrica de Médicos Sin Fronteras:

> "Cómo me indigna oír que los derechos de las patentes no constituyen una barrera al tratamiento aquí en Sudáfrica. He visto hombres y mujeres jóvenes morir víctimas de tumores cerebrales asociados al SIDA, tras padecer insoportables dolores de cabeza. He visto niños llenos de cicatrices provocadas por dermatitis asociadas al SIDA, incapaces de dormir por el dolor. Sabía que una terapia antirretroviral podía ayudarles, y que la única barrera que lo impedía era el coste del medicamento patentado" (Boulet, Garrison, & Hoen, 2003).

Por ello, viendo la crítica situación de emergencia, el presidente Nelson Mandela y la ministra de sanidad del recién elegido gobierno sudafricano, Nkosazana Zuma, decidieron establecer un comité que reformara completamente el sistema sanitario del país, que hasta entonces estaba enfocado a prestar servicios privados a los ciudadanos blancos y de rentas altas. En 1996, el comité presentó la Política Nacional de Medicamentos revisada, y en 1997 se llevaron a la legislación sudafricana mediante la Ley de Control de Medicinas y Sustancias Relacionadas. Algunas de sus medidas para luchar contra la epidemia de SIDA que estaba segando las vidas de sus ciudadanos incluían la aprobación de importaciones paralelas de genéricos (más asequibles que los patentados como sabemos) y la producción local de genéricos, para poder superar la barrera de los precios por las patentes. Esto suponía una clara amenaza para los beneficios de las *Big Pharma*, no ya porque el mercado sudafricano supusiera una parte importante de sus negocios, sino porque temían que se extendiera una ola de regulaciones similares por el resto del mundo (WHO, 2006).

La respuesta de estas empresas a través de sus portavoces no se hizo esperar: el embajador de Estados Unidos en Pretoria empezó a presionar al gobierno de Mandela para derogar la ley. Al mismo tiempo, 47 congresistas emplazaron al Representante de Comercio de los Estados Unidos para que tomara medidas contra la nueva legislación. Tras ello, el Representante de Comercio incluyó a Sudáfrica en la "lista de vigilancia" para revisar los acuerdos comerciales entre ambos países, e impuso medidas económicas contra las importaciones de productos sudafricanos. A su vez, el Departamento de Estado amenazó al gobierno del Congreso Nacional Africano con retirar todas las ayudas al desarrollo si la situación no se resolvía de manera favorable a los intereses de las farmacéuticas (Fisher & Rigamonti, 2005). En palabras del propio Departamento de Estado, las instituciones gubernamentales de los Estados Unidos llevaron a cabo "una campaña asidua y coordinada" para conseguir derogar la ley sudafricana, en la que también se involucraron las instituciones de la Unión Europea y de Suiza (USDoS, 1999).

De ahí hasta el fin del conflicto en 2001, se sucedieron acciones legales de más de 40 empresas farmacéuticas que consideraban que la nueva ley era inconsistente con las disposiciones del Acuerdo TRIPS (WHO, 2006). Sin embargo, Estados Unidos no se atrevió a presentar una demanda contra Sudáfrica ante la Organización Mundial de Comercio porque sabía que la situación de emergencia sanitaria del país era lo suficientemente relevante como para poder implementar las flexibilidades permitidas por el TRIPS. Finalmente, debido a la gran presión internacional de ONGs y de las organizaciones de enfermos de SIDA de todo el mundo, que consiguieron llevar a los grandes medios la actuación de unas compañías que anteponían sus beneficios a la vida de miles de personas, en 2001 se retiró la demanda contra el gobierno sudafricano (SwissInfo, 2001).

Se corre el riesgo de pensar que las farmacéuticas, como si de una película de Hollywood se tratara, hayan aprendido la lección y valoren más el acceso a la salud de los enfermos que los beneficios que les reportan los medicamentos patentados. Pero si vemos los datos que ofrece la Comisión de la Unión Europea (UE, 2009) para los 6 años siguientes a la controversia, pronto descubriremos que no es así. De 2001 a 2007, las farmacéuticas en la Unión Europea impusieron más de 700 demandas por violación de sus derechos de propiedad intelectual contra productores de medicamentos genéricos que ofrecían el mismo producto a un precio varias veces inferior. Estas demandas se realizaron tras el vencimiento del período de protección que ofrecía la patente, por lo que sabían que los productores genéricos tenían derecho en la mayoría de los casos para desarrollar sus fármacos. De hecho, los productores genéricos ganaron la mayoría de los juicios (contando con muchos menos recursos para pagar los gastos judiciales y abogados). No obstante, los juicios duraron una media de 3 años, con lo que lograron retrasar la entrada de genéricos (incluyendo los que no fueron demandados) una media de 7 meses. Puede parecer mucho esfuerzo para tan poco resultado, pero tenemos que tener en cuenta que un fármaco patentado medio logra un millón de dólares de ventas protegidas cada día en Estados Unidos [9] (Spinelli, 2015). La Comisión de la Unión Europea estima que gracias a las entradas de los genéricos, los gobiernos de la Unión se ahorraron más de 14 mil millones de euros en gastos sanitarios, pero que este ahorro pudo haber sido 3 mil millones de euros mayor si los genéricos no hubieran tardado esos 7 meses en conseguir entrar en el mercado europeo (UE, 2009). Con el progresivo envejecimiento de la población europea y el mayor consumo de medicamentos que observamos con el paso del tiempo (ver Mapa 4), podemos asegurar que el impacto de la entrada de medicamentos genéricos cada vez es mayor, y mayores son por consiguiente los esfuerzos de las *Big Pharma* por proteger sus productos, por ejemplo, mediante la "renovación" de sus patentes a través de los llamados medicamentos *me-too*, como luego veremos.

---

[9] Estados Unidos cuenta con una población que no llega al 65% de la de la Unión Europea, por lo que esas cifras pueden ser mayores para la UE.

Pero la lucha contra los genéricos no se limita tan sólo a los juzgados. Estas empresas, igual que en los Estados Unidos, cuentan con un gran número de *lobbyists* en la Unión Europea, además de con sistemas de puertas giratorias y otros mecanismos para influir en los Estados y en las instituciones europeas (Cronin, 2013). Con todo ello buscan no sólo moldear la legislación según sus intereses, sino también conseguir que los Estados actúen a su servicio en algunos casos: por ejemplo, requisando en los puertos los fármacos genéricos de las industrias de la India y de Brasil, hecho que se repitió hasta provocar una demanda de estos países a través de la Organización Mundial del Comercio (Lynn, 2009). Lo peor de todas estas requisas es que ni siquiera eran para evitar que entraran en el mercado común europeo, sino que se realizaron con el único objetivo de hundir a esos competidores, sin importar las vidas que dependieran del acceso a esos medicamentos, ya que las mercancías estaban en tránsito hacia países en desarrollo (Bonadio & Cantore, 2010).

Recapitulando, se ha visto cómo el sistema actual de patentes sirve los intereses de las grandes empresas farmacéuticas, otorgándoles el monopolio casi global sobre sus fármacos, lo que les garantiza unos volúmenes de ingresos protegidos abrumadores. Y para defender esos derechos, las *Big Pharma* no dudan en dirigir las instituciones públicas, ir contra la democracia, contra el derecho tanto nacional como internacional, y, en última instancia, contra los derechos humanos y la vida de las personas. ¿Cómo justifican entonces la perpetuación de este sistema tan ineficaz para mejorar la salud humana en todos los rincones del mundo?

El Informe que presentó la Comisión de la OMS para los derechos de propiedad intelectual, la innovación y la salud pública enumera varias razones: que permiten una división del trabajo más flexible y por tanto más eficiente; evita que los descubrimientos se conviertan en secretos de empresa; son una señal muy efectiva para atraer capital, especialmente en las *start-ups* tecnológicas. Pero por encima de todas ellas, siempre se destaca una razón primordial: que el monopolio que garantizan las patentes suponen un incentivo a la innovación, por lo que una mayor protección de los derechos de propiedad intelectual implican directamente una mejora de la salud de todas las personas (WHO, 2006). Veamos hasta qué punto es veraz esta afirmación.

## 3.2. Investigación y desarrollo

La industria farmacéutica afirma que su misión principal es la investigación con el objetivo de desarrollar fármacos que mejoren la salud y calidad de vida de las personas ("Our Mission | PhRMA," n.d.). Es esta su vocación y el argumento que se utiliza para defender la necesidad de las patentes y los altos precios.

Uno de los ejemplos que más se encuentran en la literatura alineada con los intereses de las empresas farmacéuticas es la del medicamento de los mil millones de dólares. Según esta línea de argumentación, los precios deben ser altos para que sea rentable investigar un nuevo medicamento, ya que el proceso de investigación y desarrollo de un nuevo producto desde que se descubre una nueva molécula hasta que el fármaco llega al mercado, supera los mil millones de dólares. De hecho, la patronal farmacéutica de Estados Unidos (PhRMA) ha aumentado recientemente esta cifra hasta los 2.600 millones de dólares estadounidenses ("Research & Development | PhRMA," n.d.). Viendo estas cifras, podríamos pensar que el sistema de patentes es necesario para incentivar una innovación que resulta tan costosa.

No obstante, los datos nos ofrecen una realidad totalmente distinta. Partamos de la advertencia que realiza la Organización Mundial para la Salud con respecto a estas cifras: en esos costes se incluyen más cosas que el coste concreto de la investigación del fármaco, desde los costes de investigaciones fallidas hasta el coste del capital (WHO, 2006). Por tanto, es evidente que hay que tener cautela con los datos que ofrece la industria farmacéutica. En primer lugar, porque nunca se ofrecen los datos brutos de los gastos en inversión de la industria, siempre se ofrecen cifras ya analizadas por "expertos independientes" (*op. cit.*).

En su seminal investigación *Demythologizing the high costs of pharmaceutical research*, Donald Light y Rebecca Warburton realizan un estudio detallado sobre este tipo de cifras. Los estudios encargados por la industria farmacéutica se basan en la información de muestras escogidas por las mismas grandes compañías que son analizadas, por lo que ya están sesgados de inicio (Light & Warburton, 2011). Pero no es sólo eso, sino que en los costes de "investigación y desarrollo" se incluyen los costes del inmobiliario que no se destina a la investigación, incluso los de administración y el gasto en renovar los equipos de producción general (PMPRB, 2014), los gastos en "contratar" a médicos con gran autoridad para que legitimen sus estudios e influyan en la opinión del resto de médicos, los gastos judiciales de la empresa para defender sus derechos de propiedad intelectual (que ya vimos que eran muy elevados) y un coste del capital del 11% (Light & Warburton, 2011). El coste del capital es un mecanismo que se utiliza para medir la rentabilidad de una inversión, es decir, cuánto obtendría la empresa si empleara el capital de una inversión en el mercado de valores con un rendimiento del 11%. Es una forma de tomar decisiones de inversión muy extendida entre las empresas, pero desde luego no supone un coste en I+D para conseguir un nuevo fármaco. Es obvio que incluir el coste del capital entre los costes de I+D puede duplicar las cifras obtenidas.

Además, las empresas toman los gastos en los que incurren para desarrollar una Nueva Entidad Molecular (NME por sus siglas en inglés), que son mucho más altos que los necesarios para mejorar una molécula ya existente y crear un medicamento más eficaz. No obstante, en la década estudiada por Light y Warburton, sólo el 35% de los nuevos fármacos registrados por la Administración de Alimentación y Medicamentos de Estados Unidos (FDA) eran NME, el resto eran moléculas ya existentes.

Finalmente, los autores también tienen en cuenta las deducciones de impuestos que logran las empresas farmacéuticas al destinar el dinero de las ventas (que iría a la cuenta de beneficios, que tributa a un tipo mayor o menor dependiendo del país y del momento) a I+D, que no sólo no tributa, sino que recibe ayudas públicas. Los autores calculan que con todo ello consiguen cubrir el 50% de los costes reales de I+D. De esta forma, teniendo en cuenta todo lo anterior, los autores cifran entre 180 y 230 millones de dólares los gastos reales para desarrollar una NME (que como ya hemos visto son la minoría de los nuevos fármacos) (Light & Warburton, 2011).

Aunque la cifra se haya reducido considerablemente con respecto a lo que publicitan las *Big Pharma*, podemos pensar que las empresas corren con grandes gastos en desarrollar nuevas medicinas que mejoren la vida de las personas, y que son las grandes contribuyentes al I+D médico. No obstante, de nuevo, la realidad desmiente estos argumentos.

Centrándonos en la industria estadounidense, la más importante del mundo, las cifras son claras. Un análisis de los fondos que contribuyeron al desarrollo de todas las NMEs que se descubrieron en Estados Unidos entre 2010 y 2016 demostró que el 100% de los fármacos había recibido financiación directa o indirecta por parte del National Institute of Health (NIH). La agencia pública aportó hasta el 80% de los fondos de I+D en algunos casos (Galkina Cleary, Beierlein, Khanuja, McNamee, & Ledley, 2018). A la misma conclusión llega otro estudio, que cifra en el 84,2% el gasto público en I+D farmacéutico sobre el total en investigación básica [10] (Burke, Light, & de Francisco, 2006). Por tanto, si la industria sólo aporta el 16% de la investigación más importante y más arriesgada (una investigadora puede tardar años en dar con un uso médico para una nueva molécula), ¿a qué destinan las empresas farmacéuticas esos exorbitantes volúmenes de I+D que publicitan? Una de las mayores expertas en este campo, Marcia Angell [11], lo tiene claro: "A diferencia de la investigación médica básica, que está financiada principalmente por el NIH, la mayoría de los ensayos clínicas están financiados por la industria farmacéutica. De hecho, es ahí donde van la mayoría de dólares de la investigación farmacéutica" [12].

---

[10] La investigación básica es el primer paso de la investigación farmacéutica y el más importante. Es el que descubre las nuevas moléculas con potencial para tratar ciertas enfermedades, y donde se analizan el comportamiento químico y la síntesis de la molécula.

[11] Marcia Angell fue editora ejecutiva durante 10 años y finalmente editora jefa de la New England Journal of Medicine, la publicación médica más prestigiosa del mundo.

Y si nos atenemos a los datos, vemos que la mayor parte de los gastos en I+D va a los ensayos clínicos: se estima que entre un 45% (Bradfield & El-Sayed, 2009) y un 60% (Terblanche, 2008) del total del I+D va a las 3 primeras fases de ensayos clínicos [13]. La cifra que nos ofrece la propia patronal se encuentra en ese rango: PhRMA lo sitúa en un 52% (PhRMA, 2016). Esto se puede entender viendo que el gasto medio en cada ensayo clínico (las 4 fases completas) fue de 30,7 millones de dólares en 2014 (Statista, 2015). ¿Por qué estos gastos tan abultados en ensayos clínicos? Existen dos razones principales: los medicamentos *me-too* (fases I-III) y el marketing de la fase IV.

## 3. 3. *Me-too* y marketing

Los medicamentos *me-too* son el resultado de pequeñas variaciones sobre fármacos ya existentes con grandes ventas (Angell, 2010). Como vimos antes, este es una de las estrategias que sigue la industria para poder expandir las patentes de sus productos estrellas: realizando mínimas variaciones sobre un fármaco, con lo que no se consigue ninguna ventaja significativa sobre el ya existente, se puede conseguir blindar la molécula durante algunos años más. De esta forma, cambios en el proceso productivo o en la composición no-activa de la medicina pueden considerarse innovaciones patentables: algunos de los medicamentos más vendidos (llamados *block-busters*) pueden llegar a estar protegidos por más de 1.300 patentes sólo en la Unión Europea (UE, 2009).

Veamos algunos datos sobre este tipo de medicamentos. De todos los medicamentos aprobados por la FDA entre 2008 y 2017, sólo un 7% aportaban alguna ventaja sobre los ya existentes (Prescrire, 2018). Esta cifra ha ido bajando con el paso del tiempo, como se puede observar en la Tabla 2. Esto confirma las hipótesis de los expertos en el tema, dado que algunos consideraban hace unos años que se estaba creando un sistema de protección intelectual que fomentaba una "pseudo innovación trivial" (Avorn, 2005) que resulta mucho más rentable que la investigación de los medicamentos necesarios para tratar las enfermedades olvidadas, como vimos anteriormente. Como resultado, sólo el 18% de los gastos en I+D van destinados al desarrollo de nuevas medicinas (NSF, 2017).

---

[12] Traducción propia. Original: *Unlike basic medical research, which is funded mainly by the National Institutes of Health (NIH), most clinical trials are funded by the pharmaceutical industry. In fact, that is where most pharmaceutical research dollars go.* (Angell, 2010).

[13] Los ensayos clínicos constan de 4 fases. Las 3 primeras son los estudios médicos en sí, la fase IV es una fase de desarrollo del producto, para ver cómo puede resultar más atractivo y venderse mejor (WHO, 2006).

**Tabla 2**. La utilidad de los nuevos fármacos con el paso del tiempo

| Años | Nuevos fármacos que suponen un avance sobre los existentes (%) |
|---|---|
| 1979-1995 | 15,6 |
| 1981-2001 | 12 |
| 2002-2011 | 8 |
| 2008-2017 | 7 |

Fuente: elaboración propia a partir de los datos de Darrow, Lexchin, & Light, 2013

Por tanto, al ofrecer tan escasos resultados, los medicamentos *me-too* tienen que ser probados en ensayos clínicos con muchos individuos para disponer de una significancia estadística relevante (Darrow et al., 2013). De ahí que los costes de los ensayos clínicos sean cada vez mayores, pero no por la exigencia de estos, sino por el número de individuos que se estudian en ellos. Con respecto al nivel de exigencia, cabe mencionar que se realizan pruebas de no-inferioridad, es decir, que el fármaco sea igual o más efectivo que el que se usa de control. Esto tendría sentido si se usasen como control los medicamentos efectivos ya en el mercado, pero en realidad se utilizan pastillas de azúcar, meros placebos, con lo que es muy fácil que un fármaco no mejor que los existentes entre en el mercado como una gran innovación sobre los anteriores (Angell, 2010). Y aunque pueda parecer que esta mala práctica sea algo poco extendido y con poca relevancia posterior para los pacientes, un estudio realizado sobre los datos de la Oficina de Revisión de los precios de los medicamentos patentados (CPMPRB por sus siglas en inglés) concluyó que del crecimiento en el gasto en compra de medicamentos entre 1996 y 2003, el 80% fue para pagar medicamentos *me-too* (Morgan et al., 2005). Y estos no son sus únicos efectos, como luego veremos.

La segunda razón que se postula por los altos precios de los ensayos clínicos son los costes de la fase IV. Como vimos anteriormente, la fase IV consiste en unos ensayos más pensados para obtener información útil para el marketing y mejorar las ventas del producto que en analizar la eficacia médica del producto. Por supuesto, aquí es donde se va la mayor parte de los gastos de los ensayos, ya que las empresas están más interesadas en que su producto se venda, sea efectivo o no, se conozcan los efectos secundarios o no. La fase 4 absorbió cerca del 40% de los gastos totales en ensayos clínicos en 2014 en Estados Unidos (Statista, 2015), y más del 13% del gasto total en I+D farmacéutico [14] (Gagnon & Lexchin, 2008). Y para dejar claro que se tratan de ensayos con finalidades publicitarias, es necesario advertir de que el 75% de estos ensayos fue realizado por el departamento de marketing de las *Big Pharma*, no el departamento de I+D (*op. cit.*).

Y es que mientras que los voceros de la industria (como PhRMA) se cansan de repetir que son el sector que más destina a I+D y que los altos precios sirven para cubrir los costes de la innovación, por lo que al fin y al cabo pagar más hoy por tus medicinas es una inversión para tener mejores tratamientos en el futuro, se destina mucho más a este tipo de propaganda que a investigar realmente. Incluso si considerásemos como ciertos los datos falseados de la industria, es algo asumido ya por toda la literatura consultada que las empresas farmacéuticas gastan más del doble en marketing y publicidad que en investigación (Forcades i Vila, 2006; Gagnon & Lexchin, 2008; Lage Dávila, 2011; Torres Domínguez, 2010; Velásquez, 2015). Por poner un ejemplo, entre 1996 y 2005 la industria farmacéutica afirmó haber gastado 288 mil millones de dólares en I+D, y 739 mil millones en marketing, unas 2 veces y media más (Hasbani & Lauzon, 2006).

La industria farmacéutica de alto valor añadido, las *Big Pharma*, que gobiernan la cadena global de valores, saben que su imperio depende de la publicidad, de hacer que el mercado piense que un medicamento de la casa Novartis siempre va a ser más fiable que uno de la casa Cipla (un productor genérico de India). ¿Cómo se articula semejante gasto publicitario de manera que las *Big Pharma* sigan manteniendo su primacía en los mercados frente a la entrada de los mucho más baratos medicamentos genéricos?

### 3.4. La economía del don y sus efectos

Las grandes empresas farmacéuticas han creado en todo el mercado relacionado con los medicamentos un sistema de compra de favores y de legitimación que algunos autores han dado en llamar 'la economía del don o del regalo' (Lessig, 2011). Ya vimos anteriormente cómo se compraba la influencia en la esfera política, veamos ahora cómo se hace en el mundo académico.

Según Angell, la Academia médica funciona como Hollywood, un mundo de estrellas y otros muchos que aspiran a serlo (Angell, 2010). Dentro de ese mundo, hay voces con mucha autoridad, capaces de influir en la opinión de todas sus colegas. Es a este tipo de personas a las que contratan con suculentas nóminas las grandes farmacéuticas para que escriban artículos "académicos" sobre sus productos, que tendrán un gran impacto en el mundo médico (Darrow et al., 2013). Además, se contratan profesionales dispuestos a maquillar sutilmente los resultados de los ensayos clínicos para que favorezcan a la empresa. De hecho, mediante meta-análisis de los artículos sobre ensayos clínicos se ha demostrado que en los ensayos financiados por las empresas hay 2,5 veces más probabilidad de que hablen a favor del producto que en aquellos realizados por institutos de salud y otras agencias independientes (Lundh, Lexchin, Mintzes, Schroll, & Bero, 2017). De esta forma, se establece una "economía del don" en la que muchos profesionales sanitarios y científicos se benefician de hacerle propaganda a las grandes empresas, sin contar el resto de innumerables maneras de comprar a las médicas a niveles más bajos: contratándoles para seminarios y conferencias, ofreciéndoles pruebas gratis, pagando los gastos de viajes y congresos… (Darrow et al., 2013). Eso sin contar con las campañas publicitarias comunes que podemos observar en cada farmacia, en la televisión e incluso en los despachos de las médicas.

---

[14] Esta cifra es claramente una subestimación porque, como vimos anteriormente, en las cifras de I+D se incluyen muchos gastos para nada relacionados con la innovación. Por ello, la cifra posiblemente sea prácticamente el doble.

Pero las *Big Pharma* no se quedan ahí, también son capaces de controlar las instituciones públicas que deben velar por la seguridad de los pacientes. En 1992, el Congreso de los Estados Unidos aprobó una ley que permitía que las empresas pagaran unas *user fees* (tasas de usuarios) a la FDA, la agencia encargada de analizar los datos que las propias empresas le remiten para la aprobación o no de un producto, de forma que se reduzca el tiempo de revisión de estos datos y el producto pueda llegar antes al mercado [15]. Lo que antes podía tardar años, con esta ley se reducía a máximo 6 meses.

Los efectos de estas medidas son considerables. Cada año, la FDA recibe en concepto de *user fees* de las empresas farmacéuticas unos 700 millones de dólares, y las *user fees* llegan a suponer hoy en día un 63% de su presupuesto (FDA, 2018). Esto, según Angell, "ha puesto a la FDA en nómina de la industria a la que regula" [16]. Con esta dependencia, se han multiplicado las aprobaciones exprés de medicamentos, sin llegar a analizar en profundidad los riesgos que los fármacos puedan tener. Esto es una gran ventaja para los *me-too*, que como hemos visto no ofrecen ventaja comparativa sobre los ya existentes, pero que pueden tener otros efectos secundarios que pasan inadvertidos por la agencia reguladora.

¿Qué consecuencias tiene esto para la salud pública? Se estima que una reducción de 10 meses en el tiempo de revisión de las agencias públicas aumenta el riesgo de sufrir una reacción adversa a un medicamento (RAM) un 18%, y los fallecimientos por esta causa aumentan en un 7,2% (Darrow et al., 2013). Hoy en día, con la corrupción institucional que observamos en la economía del don, donde los datos son maquillados y falseados por las empresas antes de ser enviados a unas agencias reguladoras financiadas por ellas mismas, se estima que 1 de cada 5 medicamentos nuevos que salen al mercado provocan RAM más o menos serias, cifra que asciende a 1 de cada 3 si el fármaco ha sido aprobado de forma exprés mediante *user fees* (Lexchin, 2012). La laxitud de las agencias reguladoras y el falseo de datos de las *Big Pharma* provoca, sólo en los Estados Unidos, unos 2'7 millones de casos de RAM (por medicamentos correctamente administrados, no por negligencia médica) y unas 130.000 muertes al año (Darrow et al., 2013; ISMP, 2017), lo que convierte a la reacción adversa a medicamentos en la cuarta causa de muerte en el país (Bonn, 1998). Las nuevas medicinas que sacan las empresas farmacéuticas al mercado no sólo son en muchos casos inferiores a las existentes, sino incluso más perjudiciales.

---

[15] Recordemos que cada día que se retrasa la entrada en el mercado de un fármaco patentado supone la pérdida de un millón de dólares en ventas protegidas.

[16] Traducción propia. Original: (…) *the user fee act put the FDA on the payroll of the industry it regulates* (Angell, 2007).

Recapitulando, hemos visto que los altos precios de los fármacos patentados no se deben a un mayor gasto en investigación y desarrollo, sino a la búsqueda de beneficios de unas empresas farmacéuticas que luchan con uñas y dientes por proteger su cuota de mercado, aunque eso conlleve un riesgo inevitable para la salud de millones de personas. Para ello, no dudan en comprar las voluntades de los legisladores y convertir las instituciones del Estado en herramienta del beneficio privado, en falsear los datos de los ensayos clínicos y los gastos en I+D, y en desarrollar medicamentos patentados con más riesgos que los que ya estaban en el mercado.

## 4. Alternativas al sistema actual

Nuestro análisis se ha articulado principalmente en dos ejes: el problema del acceso a los medicamentos en los países en desarrollo, y el papel que en ello juegan las industrias farmacéuticas de los países desarrollados. Por ello, un problema global como este requerirá de respuestas globales, diferentes según el desarrollo de cada región, pero complementarias entre sí.

### 4.1. Propuestas para los países en desarrollo

Todas las expertas en el tema lo tienen claro: el principal problema con respecto al acceso a los medicamentos en los países en desarrollo son los altos precios de las medicinas disponibles (también el poder adquisitivo de las familias, pero este problema requiere un tratamiento mucho más complejo y a largo plazo del que se puede desarrollar en un trabajo como este). Ante ello, se ha demostrado que la política económica más eficaz es el aumento de la competencia mediante la producción de genéricos (Tobar & Sánchez, 2005). Sin embargo, también hemos visto que la importación de productos genéricos encarece mucho los productos, por lo que una solución más integral sería la producción local de medicamentos genéricos. De nuevo, esto también tiene sus limitaciones, especialmente la necesidad de poder disponer de un mercado amplio para poder aumentar la producción hasta donde se alcancen economías de escala. Por todo esto, se hace imprescindible la cooperación interestatal para poder desarrollar industrias farmacéuticas regionales, donde la transferencia de conocimientos y tecnología sea un pilar fundamental.

El mejor ejemplo del que se puede aprender es la industria farmacéutica cubana. Cuba tiene uno de los sistemas de salud pública más desarrollados del mundo, teniendo 7,5 médicos por cada 1000 habitantes, 3 veces más que en Estados Unidos o Canadá, y de 4 a 10 veces más que los países con un PIB per cápita similar al cubano (CIA, 2018). Cuba, pese a su reducido tamaño y al embargo económico, no necesita importar medicamentos del extranjero: casi el 90% de los fármacos vendidos en sus farmacias se producen en el país (WHO, 2015). Por supuesto, casi la totalidad de sus medicamentos son genéricos, sólo el 15% de los productos son originales.

¿Cómo logra esto? En primer lugar, con una gran inversión en educación, que ha hecho de Cuba una de las líderes en investigación médica, especialmente en el campo de la biomedicina y la biotecnología. El presupuesto de I+D se reparte entre diferentes centros e instituciones públicas, que cooperan (reduciendo la duplicidad de gastos que se da en los sistemas privados) para lograr el objetivo de resolver los problemas de salud de la población, no el de obtener beneficio económico. En segundo lugar, hace uso de las flexibilidades de los Acuerdos TRIPS para poder producir las medicinas que necesite la población, sin atender a los derechos de propiedad intelectual de las grandes compañías farmacéuticas. En tercer lugar, Cuba transfiere su tecnología y su capital humano a otros países del Sur global (incluso la Unión Europea importa su tecnología), destacando su cooperación con otros países del ALBA, especialmente Brasil, que tiene la capacidad productiva para que la investigación realizada en Cuba se plasme en fármacos de bajos precios accesibles para los pacientes de la región de Iberoamérica (WHO, 2015).

Este sería el modelo a seguir por el resto de países en desarrollo, pero también se le reconocen limitaciones, especialmente institucionales. Prácticamente ningún Estado en desarrollo tiene la capacidad y la fuerza política del Estado cubano, y debemos recordar que sus presupuestos en salud son mucho más limitados. No obstante, esto no es ninguna limitación insuperable, es más bien un escollo político. El camino para una mejor salud pública pasa por la cooperación regional, la lucha conjunta para eliminar el servicio a la deuda ilegítima, y la inversión tanto en sanidad pública (para frenar la hemorragia de profesionales sanitarios) como en educación y capital humano, para así lograr a largo plazo un sistema de innovación al servicio de los intereses de los ciudadanos de la región, no de las empresas que hacen beneficios con la muerte.

## 4.2. Propuestas para los países desarrollados

Por otro lado, los países desarrollados deberían utilizar su mayor poder para facilitar un cambio en el sistema de protección de derechos de propiedad intelectual a nivel global. Los Estados deben demostrar su compromiso político con la salud y la vida humana frente a los intereses de las empresas farmacéuticas. Una mayor transferencia tecnológica, sumada a la inversión ética en las incipientes industrias farmacéuticas en los países del Tercer Mundo y a la cooperación académica con ellos sería un gran paso adelante para mejorar la salud de millones de personas.

En su propio territorio también deben mejorar las políticas de acceso a medicamentos genéricos, con lo que los Estados se ahorrarán miles de millones de dólares en la salud de poblaciones cada vez más envejecidas. Este ahorro podría servir para mejorar los sistemas públicos de sanidad, o para medidas sociales que luchen contra la pobreza en la era de la desigualdad.

Por supuesto, estas medidas deben ir acompañadas a mayores controles de las empresas farmacéuticas, cuando no su nacionalización [17], para evitar que se antepongan los beneficios sobre la salud de las personas. Para ello, se deben controlar los precios de los medicamentos, aumentar la seguridad de los medicamentos mediante un mejor análisis de las pruebas clínicas, y acabar con la situación de *free riding* que hacen las empresas privadas sobre las investigaciones públicas.

---

[17] La nacionalización de la industria farmacéutica en Cuba ha logrado unos fantásticos resultados en términos de esperanza de vida y acceso a los medicamentos (WHO, 2015).

# 5. Conclusiones

A lo largo de todo este trabajo, hemos analizado las dos dimensiones de la pésima asignación de recursos de investigación y de producción que lleva a cabo el mercado en el acceso a los medicamentos y la seguridad de su consumo.

Ha quedado claro que la falta de acceso a los medicamentos, especialmente en los países en vías de desarrollo, es un problema real que provoca millones de muertes cada año. Es un problema causado principalmente por la dejadez de la industria farmacéutica, que no investiga enfermedades a las que no ve un beneficio económico sustancial en las ventas potenciales, y por los altos precios a los que se enfrentan familias con un bajo poder adquisitivo. La falta de desarrollo institucional y la prioridad del pago al servicio de la deuda pública hacen que los presupuestos en sanidad sean muy bajos, y que, por consiguiente, sean las familias las que tengan que costear la mayor parte de su tratamiento médico.

Las familias tienen, así, que pagar de su propio bolsillo tratamientos que tienen unos precios altísimos, lo que supone que en muchos casos tengan que elegir entre vender sus activos y endeudarse (de por vida, en el caso de las enfermedades crónicas) o no poder acceder a las medicinas que necesitan. Esta situación tan injusta está causada en gran medida por la imposibilidad de producir los fármacos en las propias regiones donde hay falta de acceso, por lo que, aunque busquen la alternativa más barata (los medicamentos genéricos), los precios suben inevitablemente por los costes de la importación y por los *marks-up* que imponen los vendedores.

Esta situación se podría evitar mediante la producción local de medicamentos, un proyecto en el que ya están invirtiendo las organizaciones internacionales. Sin embargo, un obstáculo impide que esto se realice: los derechos de propiedad intelectual, ejemplificados en las patentes farmacéuticas. Desde finales del siglo XX, se ha establecido un marco regulador muy protector con estos derechos, con el argumento de que una mayor protección favorecerá la innovación y la investigación de nuevos medicamentos que mejoren la salud en todo el mundo. Pero, como hemos visto, esto no es que no se cumpla en todos los casos, es que es directamente mentira. Las grandes compañías farmacéuticas (*Big Pharma*) ponen todo su poder e influencia sobre los Estados al servicio de sus beneficios económicos, aunque sea para evitar que los gobiernos puedan luchar contra la epidemia de SIDA que asola muchos países del África Subsahariana, aunque priorizar sus beneficios conlleve la muerte de miles de personas.

Tanto es así, que la industria farmacéutica no se mueve por lograr la mejora de la salud de las personas, no lo considera mucho más que un mero objetivo intermedio que les permite hacerse con mayores beneficios económicos. No obstante, si pueden maximizar sus beneficios económicos sin mejorar la salud pública, las *Big Pharma* sacarán al mercado productos que no suponen ningún avance médico real con respecto a los ya existentes, para poder blindar sus medicamentos más vendidos con nuevas patentes que, como hemos visto, evitan que esos productos en las regiones donde más necesarios son. Y no importa que haya enfermedades que sean una lacra en muchos países, ni que millones de personas las sufran, y mientras no estén siendo investigadas: destinan la mayor parte de su presupuesto en publicidad para mantener sus cuotas de mercado y sus partidas de I+D en desarrollar productos prácticamente iguales que los que ya estaban en el mercado.

Si pueden maximizar sus beneficios poniendo en peligro la salud pública, no dudarán en llevar al mercado productos farmacéuticos cuyos efectos secundarios no han sido suficientemente estudiados. No dudarán en cambiar la legislación para poner en nómina a las instituciones públicas que están encargadas de controlar la seguridad de los fármacos. Si maximizar sus beneficios supone la muerte de millones de personas en las regiones más pobres por falta de acceso a los medicamentos y de cientos de miles en las zonas más ricas por el consumo de medicamentos llevados al mercado amañando los controles, bienvenidos sean estos beneficios. Esto no es un problema individual dependiendo de los valores de los ejecutivos concretos de las grandes empresas farmacéuticas. En palabras de Marx, el capitalista "no es más que capital personificado. Su alma es el alma del capital. Pero el capital tiene un solo impulso vital, el impulso de valorizarse, de crear plusvalor" (Marx, 1867: 195). Por ello, la situación no cambiará con el paso del tiempo mientras la estructura política no cambie, de forma que las empresas no tengan carta blanca para tomar decisiones que impliquen violaciones de derechos humanos.

Para hacer frente a semejante situación, hemos planteado una batería de propuestas, que podemos resumir en sus dos ramas: propuestas para los países en desarrollo y otras para los ya desarrollados. Para el primer grupo, un ejemplo claro a seguir es el caso de Cuba, pero aplicando su política de salud pública dentro de las limitaciones que impone el sistema de mercado y teniendo en cuenta la coyuntura concreta de cada región. Los pasos a seguir serían el fomento de la producción local (pero a nivel regional) de medicamentos, mediante planes comunes de educación y transferencia tecnológica entre los distintos países de cada región, buscando las economías de escala en la producción de medicamentos y un sistema de investigación centrado en curar las enfermedades endémicas de la zona. Para ello, sería necesario profundizar en las organizaciones de cooperación económica ya establecidos, como puede ser el ASEAN o el Área de Libre Comercio Africana que está pactada dentro de la Unión Africana, haciéndolas mucho más que simples organizaciones para el libre comercio.

Es imprescindible que los Estados afronten este problema global con una mentalidad global, uniéndose para luchar por la salud de sus ciudadanos y solidarizándose más con las vidas de los ciudadanos de otros países más que con los beneficios de las empresas autóctonas. El gran poder que acumula la industria farmacéutica limita la posibilidad de acción real de los Estados para hacerles frente. En el caso de África, como ya vimos, la farmacéutica estadounidense Pfizer ingresa en un año una cantidad mayor que lo que producen (por separado) el 83% de los países africanos. Por ello, un solo Estado no conseguirá mejorar sustancialmente la situación de las medicinas en su territorio, y la cooperación política para hacer un frente común y la cooperación en investigación (con perspectiva ética, no empresarial) son imprescindibles como proyecto de política de salud pública.

De la misma manera, los Estados más desarrollados, primeramente, deben no utilizar sus instituciones y su poder diplomático para defender los intereses de la industria farmacéutica. Además, deben tender a un uso más racional de los fondos públicos de sanidad, gastando esos recursos limitados en mejores servicios de salud y en medicamentos genéricos, en vez de utilizar el dinero público para comprar medicamentos originales con un mayor coste, es decir, en vez de utilizar los fondos públicos para engordar los beneficios de las empresas privadas. Un uso más eficiente de los recursos públicos permitirá mejorar la sanidad pública y otras medidas sociales.

Igualmente, todos los Estados, sea cual sea su nivel económico, deben asegurar que los medicamentos que entran en el mercado sean totalmente seguros para el consumo y que sus efectos secundarios sean razonables con respecto a los beneficios terapéuticos que suponen. Para ello, las instituciones públicas deben ser lo suficientemente independientes económica y políticamente como para poder decidir sobre la seguridad del consumo en base a criterios científicos y datos verificados.

Estas medidas pueden parecer un horizonte utópico en una situación donde las correlaciones de fuerza parecen indicar que se perpetuará a lo largo del tiempo. Pero en otro tiempo, ¿quién hubiera pensado que en muchos países los ciudadanos tengamos acceso a la sanidad pública de forma gratuita? ¿Quién hubiera pensado que alcanzaríamos los 80 años de esperanza de vida?

Ha llegado la hora de que los Estados se pregunten qué intereses van a perseguir, los de las empresas farmacéuticas o los de sus ciudadanos. Ha llegado la hora de que los ciudadanos nos preguntemos en qué clase de democracia vivimos si priman los beneficios privados a la salud pública. El problema está ya definido, ya sólo queda actuar.

# Bibliografía

# Anexo Cartográfico

**Mapa 1**. Proporción del gasto en medicamentos sobre
el gasto total de las familias en salud en África, 1995-2014 (%)

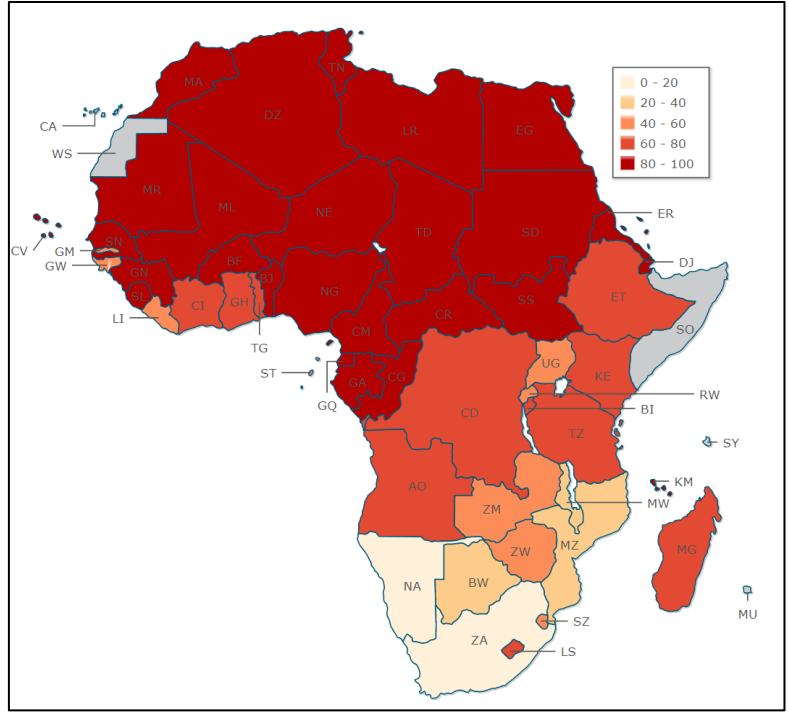

Fuente: elaboración propia a partir de WB, 2018.

**Mapa 2.** Número de patentes desde la aprobación de los Acuerdos TRIPS (1995)

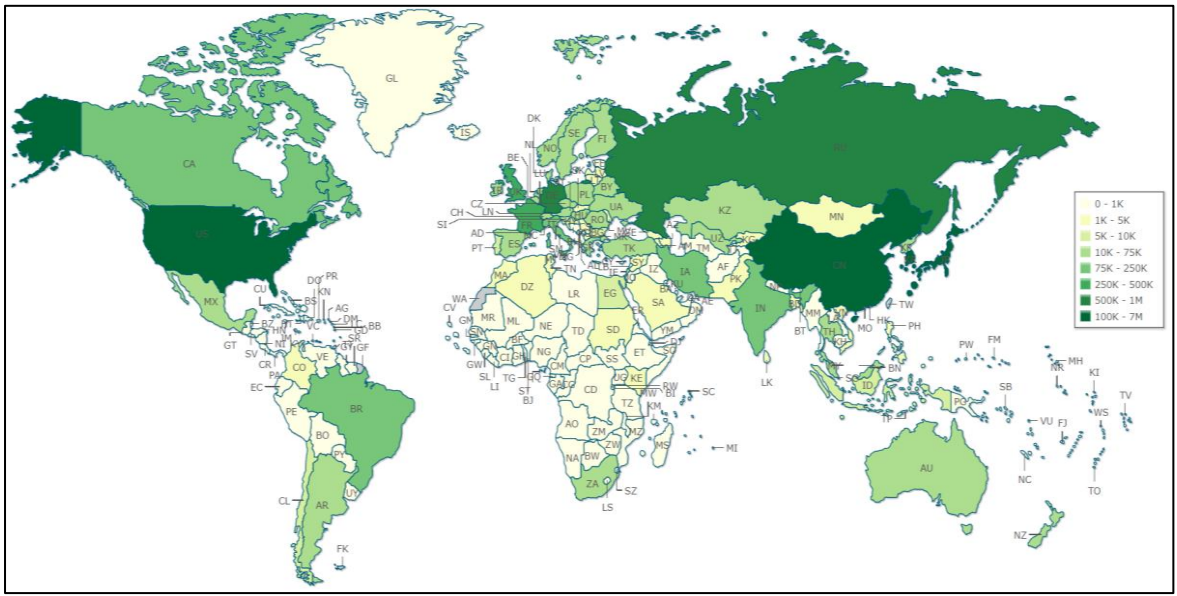

Fuente: elaboración propia a partir de WB, 2018.

**Mapa 3**. PIB de África vs. Ingresos de Pfizer (año 2016)

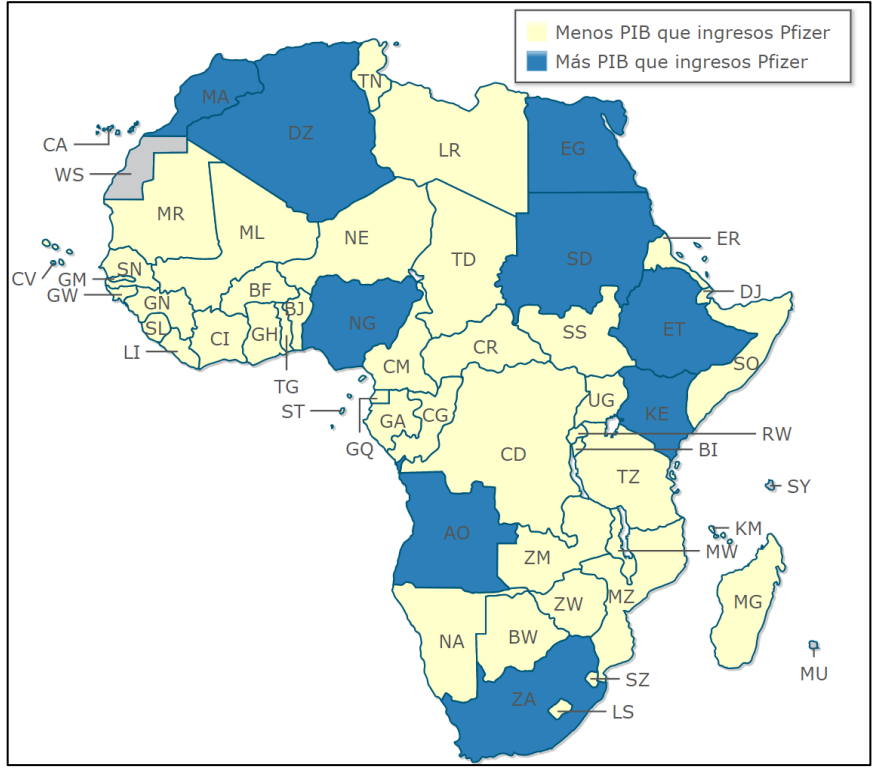

Fuente: Elaboración propia a partir de WB, 2018 y Statista, 2018

**Mapa 4.** Gasto per cápita en medicamentos en 2016 ($)

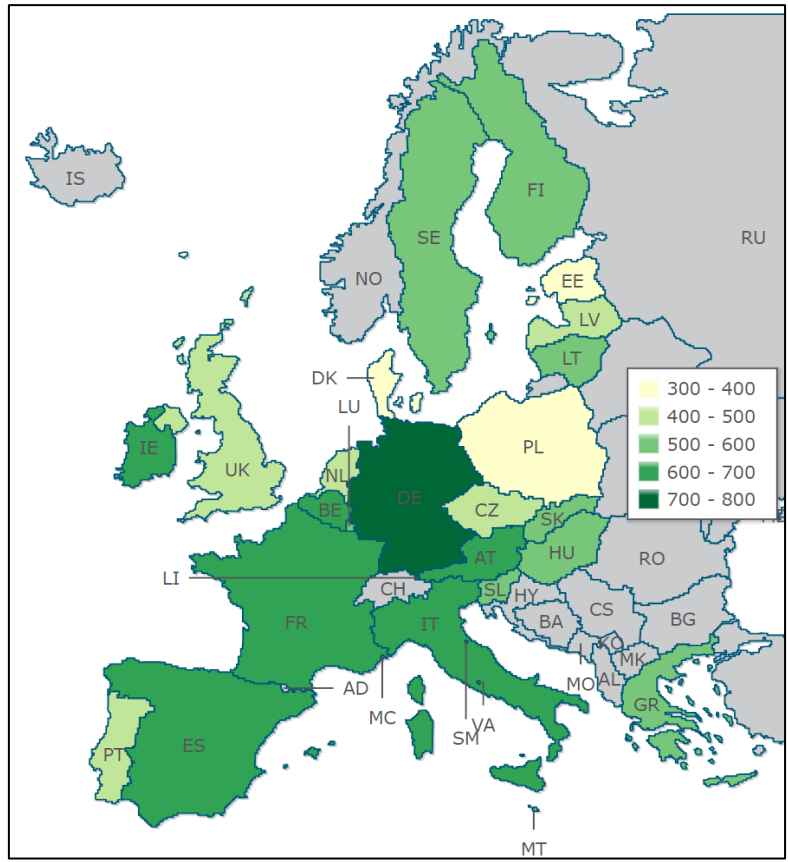

Fuente: elaboración propia a partir de datos de OCDE, 2018

**Mapa 5.** Proporción del gasto en medicamentos sobre
el gasto total de las familias en salud en Sudamérica, 1995-2014 (%)

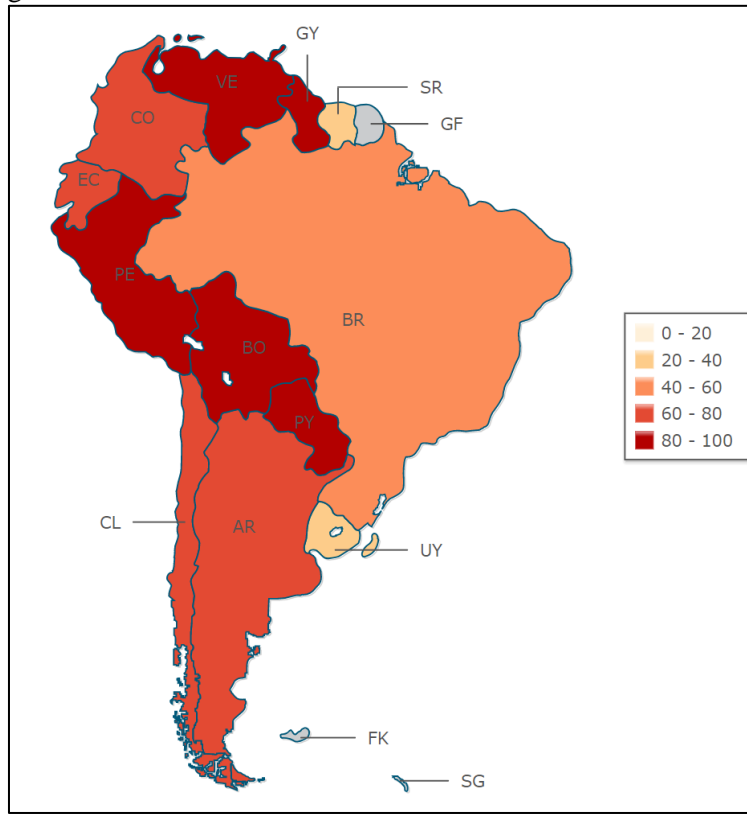

Fuente: elaboración propia a partir de WB, 2018.